\documentstyle[12pt]{article}
\def\be{\begin{equation}}
\def\ee{\end{equation}}
\def\ba{\begin{array}}
\def\ea{\end{array}}
\def\bea{\begin{eqnarray}}
\def\eea{\end{eqnarray}}

\begin{document}
\begin{center}
{\Large \bf The formation and decay of superheavy nuclei produced in
$^{48}Ca$-induced reactions} \\

\bigskip

{\bf Sushil Kumar$^{a}$, M. Balasubramaniam$^{a,b}$, Raj K. Gupta$^{a,b,c}$,\\
G. M\"unzenberg$^c$ and W. Scheid$^b$}

\medskip 

$^a$ Physics Department, Panjab University, Chandigarh- 160014, India\\
$^b$ Institut f\"ur Theoretische Physik, J-L Universit\"at, 35392 Giessen, 
Germany\\
$^c$ Gesellschaft f\"ur Schwerionenforschung mbH, 64220 Darmstadt, Germany\\
\medskip 

\end{center}

\vspace*{0.3cm}

\begin{center}
{\bf Abstract}
\end{center} 

\medskip

\baselineskip 18pt
The formation of superheavy nuclei in $^{48}Ca$+$^{232}Th$, $^{238}U$, 
$^{242,244}Pu$ and $^{248}Cm$ reactions and their subsequent decay are 
studied within the quantum mechanical fragmentation theory (QMFT) and the 
QMFT based preformed cluster-decay model (PCM) of Gupta and collaborators.
According to QMFT, all these $^{48}Ca$-induced reactions are cold fusion
reactions with relative excitation energies larger than for the $Pb$-induced
cold fusion reactions and smaller than for the lighter beam i.e. $Mg$, $Si$
or $S$-induced hot fusion reactions. The same reactions were first suggested
by Gupta et al. in 1977 on the basis of QMFT, and this study re-establishes 
the same result. In fact, for such heavy isotopes of Z=110 to 116, $^{50}Ca$ 
is shown to be a better beam for cold fusion, but $^{50}Ca$ is a radioactive 
nucleus. The $\alpha$-decay half-lives of these nuclei after 3n and/ or 4n 
evaporations, i.e. of the evaporation residues of these compound systems, 
calculated on PCM compare reasonably well with experiments published by 
Dubna group and another recent calculation. As expected for such rare 
decays, PCM calculations show that the $\alpha$-preformation factors are 
small, $\sim 10^{-8}$ to $10^{-10}$. The possible competition of 
$\alpha$-decays with heavy cluster emissions from these superheavy nuclei 
is also probed from the point of view of searching for new nuclear structure 
information and possible future experiments with such exotic nuclei. The 
decay half-lives for some clusters are in fact shown to be lower than the 
limits of experiments for nuclei with enough available atoms.  

\vfill\eject

\section{Introduction}
The formation of new and superheavy elements through the use of highly 
neutron rich beam of $^{48}Ca$ on neutron rich actinide targets, such as 
$^{244}Pu$, $^{248}Cm$ and $^{252}Cf$, was first suggested by Flerov 
\cite{flerov69} and was used at his JINR Dubna U-300 heavy-ion cyclotron 
for the successful synthesis of $^{252}No$ in a cold (17-18 MeV excitation) 
2n emission reaction with $^{206}Pb$ target \cite{flerov76}. However, the 
same could not be continued to heavier elements due to the poor sensitivity 
level of the equipment at that time \cite{oganess78}. Theoretically, Nix 
\cite{nix73} later suggested the use of very asymmetric target-projectile 
combinations, such as $^{250}Cm$ and $^{257}Fm$ with $^{48}Ca$, on the basis 
of the kinetic energy calculations made for different nuclear shape 
configurations by using an idealised liquid drop model. In a more complete 
theory, called quantum mechanical fragmentation theory (QMFT)
\cite{fink74,maruhn74,gupta75,sandu76,gupta77,gupta77a,gupta77b,gupta77c},
based on two centre shell model and Strutinsky renormalization procedure,
Gupta, S\v andulescu and Greiner \cite{gupta77b} also suggested the use of
$^{48}Ca$ beam, on the minimum energy considerations, with even-even
$^{196-204}Hg$, $^{204-208}Pb$, $^{232}Th$, $^{234-238}U$, $^{240-244}Pu$ 
and $^{244-248}Cm$ targets for synthesizing the various isotopes of even 
Z=100, 102, and 110-116 (only even Z nuclei were studied). The calculations 
also showed that for the heavier mass isotopes at the minimum in energy 
either the projectile is not $^{48}Ca$ or the corresponding target nucleus 
is not stable in nature. On the other hand for more neutron-deficient 
isotopes, the potential energy surfaces tend to become flat and $^{48}Ca$ 
is no more a favoured nucleus since other new minima start to develop. 
Furthermore, for Z=104-108, the targets to be used with $^{48}Ca$ projectile 
are the unstable, very difficult to handle $Po$, $Rn$ and $Ra$ nuclei. The 
key consideration behind the QMFT is the {\it shell closure effects} of one 
or both the reaction partners and hence of "cold reaction valleys" or the 
reaction valleys leading to {\it cold} fusion. Four or five such cold 
reaction valleys were always found to exist, $^{48}Ca$ being one of them. 
Their relative excitations would mean the cases of "cold", "warm/tepid" and 
"hot" fusion reactions. In other words, all cases of potential energy minima 
are in fact of "cold" reactions with different relative excitation energies, 
and the real "hot" reactions are the ones coming from outside the potential 
energy minima.

More recently, an intense $^{48}Ca$ beam at low consumption of material in 
the ion source was developed \cite{kutner98} and used with the upgraded 
Dubna cyclotron U-400 and new recoil mass separators (VASSILISSA or GNS) for 
the formation of superheavy elements 110 to 116
\cite{oganess00,oganess99,oganess99a,oganess00a,oganess99b,oganess00b}.
The targets used are $^{232}Th$, $^{238}U$, $^{242,244}Pu$ and $^{248}Cm$,
and, for near the Coulomb barrier energies, the excitation energy 
$E^*\sim$30-35 MeV, in between the one for cold ($E^*\sim$10-20 MeV) and hot 
($E^*\sim$40-50 MeV) fusion reactions. Thus, the $^{48}Ca$-induced reactions 
are referred to as "warm/ tepid" fusion reactions, in between the "cold" and
"hot" fusion reactions. For some $^{48}Ca$-induced reactions, $E^*\sim$40 
MeV also. Thus, the resulting compound systems de-exite by 3n and/ or 4n 
evaporations, compared to 1n and 2n in cold and 5n in hot fusion reactions. 
The cold fusion reactions are based on $Pb$ and $Bi$ targets with $Ti$ to 
$Zn$ beams, and the hot fusion reactions use heavy actinide targets from 
$Th$ to $Cf$ with very light $C$ to $S$ projectiles. In all the "warm" 
fusion reactions studied here, the final nuclei formed after the 
evaporation of neutrons (and $\gamma$-ray emission), the evaporation 
residues EVRs, are relatively long-lived nuclei and decay only via 
$\alpha$-particles, ending in spontaneous fission (SF) of the last nucleus 
at the edge of the stability region. This is the complete fusion reaction 
channel, giving the $\alpha$-genetically related nuclei, called 
$\alpha$-decay chain. Another decay channel for the excited compound system 
is that of fission-like decays: the fusion-fission, quasi-fission, etc., 
which has also been of interest both experimentally 
\cite{kratz74,bock82,toke85,itkis02} and theoretically 
\cite{gupta77d,aroumou87,gupta97}. It may be relevant to mention here that 
the fusion-fission process in hot superheavy nuclei could also be treated 
as a dynamical cluster decay process \cite{gupta02} advanced recently for 
light and medium mass hot compound systems.

In this paper, we first re-investigate the use of the $^{48}Ca$ beam for the
synthesis of superheavy elements within the dynamical fragmentation theory
via cold fusion reactions. This has become essential because the binding 
energy data has improved considerably since its last use \cite{gupta77b} by 
some of us in 1977. Specifically, we consider the compound systems 
$^{280}110^*$, $^{286}112^*$, $^{290,292}114^*$ and $^{296}116^*$, formed in 
reactions of $^{48}Ca$ beam on $^{232}Th$, $^{238}U$, $^{242,244}Pu$ and 
$^{248}Cm$ targets at various incident energies. As already stated above, 
the excitation energy $E^*=$ 32-35 MeV in these reactions and hence they all 
de-excite by 3n and/ or 4n evaporations, to give $^{277}110$, $^{283}112$, 
$^{287,288,289}114$ and $^{292}116$ parents. Then, as a second aim of this 
paper, we investigate the observed $\alpha$-decay characteristics 
of these nuclei within the preformed cluster decay model (PCM) of Gupta and 
Collaborators \cite{gupta88,malik89,gupta91,kumar94,gupta94,gupta99} which
is also based on the dynamical fragmentation theory. Furthermore, a possible 
branching of $\alpha$-decay to some (theoretically) most probable heavy 
cluster decays is also studied with a view to look for some new or known 
magic daughters in this exotic process of, so-called, cluster radioactivity. 

The paper is organised as follows. A brief description of the fragmentation 
theory and PCM is given in section 2. The results of our calculations, 
compared with Dubna experiments and other recent works, are presented in 
section 3. A summary of our results is added in section 4.

\section{The Theory}
\subsection{Quantum mechanical fragmentation theory}
The quantum mechanical fragmentation theory (QMFT) is a dynamical theory of
heavy-ion collisions, specifically the three cold processes of fusion, 
fission and cluster radioactivity, worked out in terms of the coordinates of
mass (and charge) asymmetry $\eta=(A_1-A_2)/A$ (and $\eta_Z=(Z_1-Z_2)/Z$),
the relative separation distance $R$, the deformations $\beta_1$ and 
$\beta_2$ of two nuclei (or, in general, fragments), and the neck parameter 
$\epsilon$. Taking motions in $\eta$ and $\eta_Z$ as weakly coupled, the 
time-dependent Schr\"odinger equation in $\eta$, 
\begin{equation} 
\label{eq:1}
H\Psi(\eta,t) = i\hbar \frac{\partial }{\partial t} \Psi(\eta,t),
\end{equation}
is solved for R(t) treated classically and the other coordinates $\beta_1$, 
$\beta_2$ and $\epsilon$ fixed by minimizing the collective potential in 
these coordinates \cite{gupta80,yamaji77}. We find that for a given 
compound system (formed by different target + projectile combinations), a 
few nucleon to a large mass transfer occurs for the target + projectile 
combinations coming from {\it outside} the potential energy minima (the real 
"hot" combinations), whereas the same is zero (no transfer at all) for the 
target + projectile combinations referring to potential energy minima (the 
"cold" combinations); see Fig. 14 in \cite{yamaji77}. This means that for 
cold reaction partners, the two nuclei stick together and form a deformed 
compound system, as is illustrated in Fig. 2 of \cite{gupta77}. On the other 
hand, a few nucleon transfer may occur if only a "conditional" saddle is 
formed (see Figs. 1 and 2 in \cite{malhotra86}). Eq. (\ref{eq:1}) is solved 
for only a small number of heavy systems, since its solution is very much 
computer-time consuming. However, certain simplifications, discussed in the 
following, have resulted from actual calculations performed over the years                       
which seem to work rather nicely.

The dynamical QMFT establishes the best target + projectile combination, 
with the variation of relative separation coordinate R allowing to pass 
continuously from a separated pair of nuclei to a cool compound nucleus 
\cite{sandu76,gupta77} or any other process like fission or deep inelastic 
collision, etc., \cite{aroumou87,gupta97}. However, the potentials 
V(R,$\eta$) and V(R,$\eta_Z$), calculated within the Strutinsky 
renormalization procedure (V=V$_{LDM}$ + $\delta$U) by using the appropriate 
liquid drop model for $V_{LDM}$ \cite{myers67} and the asymmetric two-center 
shell model (ATCSM) of Maruhn and Greiner \cite{maruhn72} for the shell 
effects $\delta$U, show that the motions in both $\eta$ and $\eta_Z$ are 
much faster than the R-motion. This means that both the potentials 
V(R,$\eta$) and V(R,$\eta_Z$) are {\it nearly} independent of the 
R-coordinate (see e.g. Fig. 1 in \cite{sandu76} or \cite{gupta77d}) and 
hence R can be taken as a time-independent parameter. Then, the 
time-dependent Schr\"odinger equation (\ref{eq:1}) in $\eta$ reduces to the 
stationary Schr\"odinger equation in $\eta$,
\begin{equation} 
\label{eq:2}
\{ - \frac{\hbar^2}{2\sqrt{ B_{\eta\eta} }} \frac{\partial}{\partial\eta}
\frac{1}{\sqrt{ B\eta\eta }} \frac{\partial}{\partial\eta} +
V_R(\eta)\} \Psi_R^{(\nu)}(\eta) = E_R^{(\nu)}\Psi_R^{(\nu)}(\eta),
\end{equation}
where R is fixed at the post-saddle point, justified by many good fits to
both fission and heavy-ion collision data \cite{gupta99a} and by an 
explicit, analytical solution of time-dependent Schr\"odinger equation in 
$\eta_Z$ coordinate~\cite{saroha83}. An interesting result of these 
calculations is that the mass (and charge) distribution yields, 
\be
Y(A_i)=\mid \psi _R(\eta (A_i))\mid ^2 {\sqrt {B_{\eta \eta}}} {2\over A},
\qquad {\hbox{(i=1,2)}}
\label{eq:3}
\ee
(and $Y(\eta_Z)$) are nearly insensitive to the detailed structure of
the cranking masses B$_{\eta\eta}$, calculated consistently by using ATCSM.
The nuclear temperature effects are included here through a Boltzmann-like 
function
\be
\mid \psi _R\mid ^2=\sum _{\nu =0}^{\infty}\mid \psi _R^{(\nu )}
\mid ^2 exp (-E_R^{(\nu )}/T), 
\label{eq:4}
\ee
with the temperature $T$ (in MeV) defined, via the compound nucleus 
excitation energy, as
\be
E^*=E_{cm}+Q_{in}={A\over 9}{T}^2-T.
\label{eq:4a}
\ee
$Q_{in}$ is the Q-value for reaction partners. Also, the shell corrections 
$\delta U$ are T-dependent, but this could not be included here since we
are using the experimental binding energies where $\delta U$ are there in
them but could not be separated out in a model independent way. However, the
T-values involved here are small, $\sim$1 MeV only. 

Thus, the static potentials V($\eta$) (and V($\eta_Z$)) contain all the 
important information of a colliding or fissioning system. Furthermore, 
since these potentials are nearly independent of the choice of R-value (as
discussed above), they are calculated at some critical distance 
$R=C_t=C_1+C_2$ where the two nuclei come in close contact with each other. 
This means fixing the neck parameter $\epsilon$=1 and the potential 
V($\eta$, $\eta_Z$) given simply as the sum of the ground state binding 
energies $B_i$ of two nuclei, the Coulomb interaction $E_c$ between them 
plus the additional attraction due to nuclear proximity $V_P$ 
\cite{blocki77},
\begin{equation}
\label{eq:5}
V(C_t, \eta,\eta_Z) = -\sum_{i=1}^{2} B(A_{i}, Z_{i}, \beta _{i}) +
\frac{Z_{1} Z_{2} e^{2}}{C_t} + V_{P}.
\end{equation}
The $C_i$, defining $C_t$, are the S\"ussmann central radii 
$C_i=R_i-({1/R_i})$, with the radii 
$R_i=1.28A_i^{1/3}-0.76+0.8A_i^{-1/3} fm$. The binding energies $B_i$ are 
from the 1995 experimental compilation of Audi and Wapstra \cite{audi95} 
and from 1995 calculations of M\"oller et al. \cite{moller95} whenever not 
available in \cite{audi95}. The charges Z$_1$ and Z$_2$ are 
determined by minimizing the potential in $\eta_Z$ coordinate, which 
automatically minimizes the $\beta_i$ coordinates. Note that the minimized 
$\beta_i$'s are {\it not} always for the spherical nuclei since the total 
binding energy $B_1+B_2$ of the reaction partners is minimized and not their 
individual $B_1$ or $B_2$. Also, it may be pointed out that $V_P$, the 
additional attraction between nuclear surfaces, was not added in our earlier 
calculations, whose effect is to change the relative heights of the 
potential energy minima, and hence the relative excitations, but not their 
positions \cite{malhotra86}. The positions of the minima are due to shell 
effects only. Also, the role of deformation in both $E_c$ and $V_P$ is 
neglected here because their combined effect is shown \cite{kumar97} to 
lower the interaction barriers but not the relative formation yields. In 
other words, Eq. (\ref{eq:5}), without V$_P$, formed the basis of our first 
calculation on "cold fusion" reaction valleys 
\cite{sandu76,gupta77a,gupta77b,gupta77c}, which was later optimized by 
adding the requirements of smallest interaction barrier, largest 
interaction radius and non--necked (no saddle) nuclear shapes 
\cite{gupta77}. Like necked-in shapes are known \cite{maruhn74,gupta75} to 
witness the preformation of fission fragments, non-necked shapes are the 
signatures of cold fusion of two nuclei \cite{gupta77}. The contribution 
due to V$_P$ is, however, added in the present calculations.

\subsection{Preformed cluster-decay model}
The preformed cluster-decay model (PCM) is based on the QMFT and hence uses 
the same coordinates as are introduced above. In a PCM, the decay constant 
$\lambda$, or the decay half-life $T_{1/2}$, is defined as
\be
\lambda ={{{ln 2}\over {T_{1/2}}}}=P_0\nu _0 P,
\label{eq:6}
\ee
with $P_0$ as the cluster (and daughter) preformation probability in the
ground state of nucleus and P the barrier penetrability which refer, 
respectively, to the $\eta$ and R motions.
The $\nu _0$ is the barrier assault frequency. In principle, the coordinates 
$\eta$ and R are coupled, but Eq. (\ref{eq:6}) is always written in a 
decoupled approximation, which further justifies our simplifying conditions 
of the last subsection.

The $P_0$ are the solutions (\ref{eq:3}) of the stationary Schr\"odinger 
equation (\ref{eq:2}) in $\eta$ for the ground-state $\nu =0$, i.e. 
$P_0=Y^{0}(\eta)$. The mass parameters $B_{\eta \eta}(\eta )$ are the 
classical hydrodynamical masses of Kr\"oger and Scheid \cite{kroeger80} 
since the two masses (cranking and hydrodynamical) are shown to give 
similar results for heavier nuclei.

The P is the WKB tunnelling probability, calculated for the tunnelling path 
shown in Fig. 1, as  $P=P_i P_b$ with
\be
P_i=exp[-{2\over \hbar}{{\int }_{R_a}^{R_i}\{ 2\mu [V(R)-V(R_i)]\}
^{1/2} dR}]
\label{eq:7}
\ee
\be
P_b=exp[-{2\over \hbar}{{\int }_{R_i}^{R_b}\{ 2\mu [V(R)-Q]\} ^{1/2} dR}],
\label{eq:8}
\ee
with the first turning point $R_a=C_t$ and the second turning point $R_b$ 
defined by $V(R_b)=Q$-value for the ground-state ($\alpha$ or cluster) 
decay. These integrals are solved analytically \cite{malik89}. This choice 
of $R_a=C_t(=C_1+C_2)$, instead of $R=R_0$, the compound nucleus radius, 
assimilates to a good extent the effects of both the deformations of two 
fragments and neck formation between them \cite{kumar97}. In other words, 
the two-centre nuclear shape is simulated here through a neck-length 
parameter which for actinides is nearly zero \cite{kumar97}. We have taken 
it to be the same for superheavy nuclei. The role of deformation in V(R) is 
to lower the interaction barriers  \cite{kumar97}, which is achieved here by 
raising $R_a$ from $R_0$ to $C_t$.

The assault frequency $\nu _0$ in (\ref{eq:6}) is given simply as
\be
\nu _0={{(2E_2/\mu )^{1/2}}\over {R_0}},
\label{eq:9}
\ee
with $E_2=(A_1/A) Q$, the kinetic energy of the lighter fragment, for the
$Q$-value shared between the two products as inverse of their masses. $\mu$
is the reduced mass here. Eq. (\ref{eq:9}) usually results in  
$\nu _0\approx 2.7\times 10^{21} s^{-1}$, whereas the more often used value 
in literature is higher for even parents and lower for odd parents 
\cite{gupta94}. We choose to use here 
$\nu _0=2.7\times 10^{21} s^{-1}$ for odd parents and 
$\nu _0=2.7\times 10^{23} s^{-1}$ for even parents.

\section{Calculations}
{\it The superheavy nucleus formation process:}
Figure 2 shows our calculated fragmentation potentials $V(\eta )$ at a
fixed $R=C_t$-value, based on Eq. (\ref{eq:5}), for some excited compound
systems $^{296}116^*$, $^{292}114^*$, $^{286}112^*$ and $^{280}110^*$ 
formed via $^{48}Ca$-induced reactions. We find in Fig. 2 that in each case 
there is a minimum corresponding to $^{50}Ca$ nucleus (the $^{48}Ca$ 
minimum lies in its immediate neighbourhood but at somewhat higher energy), 
which is in addition to other minima at symmetric (or nearly symmetric), 
the $^{208}Pb$ and $^{86}Kr$ or their neighbouring nuclei, respectively, 
with $Z=80\pm 2$ and $36\pm 2$ (the $Pb$ and $Kr$ minima merge into one 
minimum for $Z\ge 116$ elements) and the ones corresponding to the 
super-asymmetric combinations using heavier actinides with much lighter 
beams. The $^{48}Ca$ minimum is deepest only for the lighter isotopes 
of these nuclei, e.g. for $^{290}116$, $^{284}114$, $^{278}112$ and 
$^{270}110$ (see Fig. 1 and Table 1 in \cite{gupta01}). Apparently, all 
these minima correspond to at least one closed shell nucleus and hence to 
"cold" reaction valleys. Also, the relative excitation energies are 
different, the $^{48}Ca$ minimum always lying higher than the $^{208}Pb$ 
(and/or $^{86}Kr$) and the symmetric (or near symmetric) minima, but deeper 
than the ones corresponding to many heavier actinides with lighter beams of
$Mg$, $Si$, $S$, etc.,. The excitation energy can be determined by the hight 
of the minimum with respect to the ground state of the compound system. 
Knowing that temperature effects are not added here in Fig. 2, the 
approximate excitation energy $E^*\sim$30 MeV for $Ca$ minima, to be 
compared with $E^*\sim$10 MeV for $Pb$ minima and $\sim$40 MeV for heavy 
actinides. For lighter elements (Z=102 and 104), the $^{48}Ca$ minimum is
deeper than the $Pb$ and $Kr$ minima, and its relative depth is only 
15 to 20 MeV \cite{gupta77b,gupta01}. This means to say that $Ca$-induced 
reactions for heavy elements ($Z>106$) give rise to "cold" compound systems 
with an "intermediate" amount of excitation energy, and hence the name 
"warm or tepid" with respect to "cold and hot" fusion. The number of 
neutrons emitted in $^{48}Ca$-induced reactions (3n or 4n) also lie in 
between the cold (1n or 2n) and hot (5n) fusion reactions. In other words, 
according to the QMFT, the $^{48}Ca$ beams are as good as the $^{208}Pb$ 
beams (or the lighter ones like $Mg$, $Si$ and $S$) for forming cold 
compound nuclei with relatively higher (or lower) excitation energies. The 
calculated interaction barriers and nuclear shapes also support that 
$Ca$-induced reactions are as good as the $Pb$-induced ones 
\cite{gupta77,gupta97,gupta01}. Finally,
it may be relevent to note that the $^{50}Ca$ nucleus, preferred for cold
compound nucleus formation in heavier isotopes of superheavy nuclei studied 
here, is radioactive whereas the $^{48}Ca$ is not a radioactive nucleus. The 
predicted use of radioactive nuclei (as a beam and/ or a target) for heavier 
isotopes of superheavy nuclei is a general result of the QMFT 
\cite{gupta01}.

We have also calculated the fragment mass distribution yields $Y(\eta )$, 
using Eq. (\ref{eq:3}), taking the view that, since fragments related to the 
minimum in the potential $V(\eta)$ are more probable, the yields $Y(\eta )$ 
must give the intermediate (two) fragment formation yields or, in short, the 
formation yields for a cool compound nucleus \cite{gupta95}, where the 
contribution of barrier penetration is not included. Fig. 3 illustrates 
our calculation for two compound systems $^{296}116^*$ and $^{292}114^*$. 
For the mass parameters, we have used the classical hydrodynamical masses 
\cite{kroeger80}. Excluding the cases of reactions involving symmetric 
combinations, in view of their forming necked-in shapes, the formation yields 
for $^{48}Ca$-induced (here $^{50}Ca$) cold fusion reactions are apparently 
larger than for the light mass beams but smaller than for $Pb$-induced 
reactions. The contribution of penetrability P would, however, quantify
these results.

{\it The $\alpha$-decay process in superheavy nuclei:}
Table 1 and Fig. 4 show the results of our calculation for $\alpha$-decay 
half-lives, in terms of $log_{10}T^{\alpha}_{1/2}$ (s), for the measured
complete decay chains of $^{277}110$, $^{287,288,289}114$ and $^{292}116$
parents (no $\alpha$-decay of $^{283}112$ parent is observed), compared with 
the experimental data published by Dubna group and a very recent calculation 
of Royer and Gherghescu \cite{royer02,royer00} based on the generalized 
liquid drop model (GLDM). The Dubna data is for $\alpha$-particle energies
$E_{\alpha}$ and decay times $\tau$, which agree with $Q_{\alpha}$-values
(within less than 150KeV) and $\alpha$-decay half-lives $T^{\alpha}_{1/2}$ 
(within less than one order of magnitude), respectively, via the well known
formula of Viola and Seaborg \cite{viola66} whose parameters are fitted to
measured $Q_{\alpha}$ and $T^{\alpha}_{1/2}$ values of 58 even-even $Z>82$,
$N>126$ nuclei \cite{sobic99}. We have, therefore, used this data to compare
with the calculated $Q_{\alpha}$ and $T^{\alpha}_{1/2}$-values. The resulting
differences are of 0.2-0.4 atomic mass numbers (compare the resulting 
$114.4^{+1.6}_{-0.8}$ and $110.2^{+1.5}_{-0.8}$ \cite{oganess00},respectively, 
with A=114 and 110). Wherever more than one chain is observed, we 
present the data for all chains. Table 1 also lists the respective 
$Q_{\alpha}$-values and the preformation factor $P_0$ and penetrability P 
calculated on PCM. The PCM calculated Q-values ($Q_{\alpha}^{cal.}$) are 
based on experimental \cite{audi95} and M\"oller et al. \cite{moller95} 
binding energies, whereas that of GLDM on Thomas-Fermi (TF) model 
\cite{myers96}. The $Q_{\alpha}^{expt.}$ are taken to be the experimental 
$E_{\alpha}$-values measured in respective $\alpha$-decay chains. In order to 
illustrate the role of Q-value, we have also calculated the decay half-lives 
within PCM using the experimental $Q_{\alpha}^{expt.}$-values, and compared with 
the results of our calculations using $Q_{\alpha}^{cal.}$. 

We first notice in Table 1 that a decrease in Q-value by about 0.5 MeV 
increases the $T_{1\over 2}$-value by about one order. This is also 
illustrated in Fig. 4 for $^{277}110$ and $^{288,289}114$ decay chains. 
Apparently, in one case (Z=114 decay chains) use of experimental Q-values 
improve the comparisons, whereas in the other case ($^{277}110$ decay chain) 
this spoils it. In general, the comparisons of $T_{1\over 2}$-values for 
the two models with experiments are within experimental errors, i.e. within 
less than two orders of magnitude. In some isolated cases, like for 
$^{269}106$ nucleus in $^{277}110$ decay chain, the trends of experiments 
are different from calculations and hence need further checking. In total, 
the data obtained do not show any strange behaviour and hence do not call 
for any special criticism \cite{armbrus00}.

The interesting result to note in Fig. 4 is that both the GLDM and PCM 
calculations give similar trends, which is due to their used Q-values.
This is illustrated in Fig. 5, where the calculated Q-values on two models 
are compared with experiments. The experimental $E_{\alpha}$-values (taken as 
$Q_{\alpha}^{expt.}$-values) are in general higher and hence are the cause of 
one to two orders of magnitude deviations between the calculated $T_{1\over 2}$
and measured $\tau$-values.

Finally, in Table 1, we have a look at $P_0$ and P, calculated on PCM. The
$P_0$ are of the order of $10^{-8}$ to $10^{-10}$ which means that all these
$\alpha$-decays are rare in nature, as is actually the case. In GLDM, 
$P_0=1$, which means that the P calculated on PCM should be higher, due to
either the higher Q-values or lower barriers. However, Fig. 5 shows that the
Q-values used in PCM are lower than those used in GLDM. Also, it may be 
noted that $\nu_0=10^{20} s^{-1}$ in GLDM, an order of one and three lower 
from that used in PCM, respectively, for odd and even parents. All these 
factors have a combined effect on the nature of predictions made by a 
particular model.  

{\it The heavy cluster-decay process in superheavy nuclei:}
Figure 6 shows the results of our calculations for some heavy cluster 
decays whose preformation factors $P_0$ are the largest and hence refer to 
the minima in fragmentation potentials $V(\eta )$, for each of the parents 
in the $\alpha$-decay chain of, say, $^{277}$110 nucleus. The heaviest 
cluster included here in the calculations is with Z=20 ($^{49-51}$Ca) 
because an earlier calculation on PCM shows that the two processes of cold 
fission and cluster decay become indistinguishable for clusters heavier 
than of mass $A_2\approx 48$ \cite{kumar94}. The results of other decay 
chains are not presented hence since this study at present is more of an
academic interest than for comparisons with experiments.

First of all we notice in Fig. 6 that, almost independent of the parent 
mass, the Q-value increases as the size of cluster increases. On the other
hand, $P$ and $P_0$ are further smaller than for $\alpha$-decay, but present 
an interesting result. We notice that though $^{10}$Be is best preformed in 
all parents, its P-values are very small. Such a result is important since 
$T_{1/2}$ is a combined effect of both $P_0$ and P ($\nu _0$ being constant). 
For $^{10}$Be, the $T_{1/2}$ is much larger than for other clusters such 
that the studied parents could be said stable against $^{10}$Be decay. In 
other words, in addition to $\alpha$-decay and fission, $^{14}$C, $^{34}$Si 
and/ or $Ca$ clusters present the best possible cases of cluster decays for
any of the parents of $^{277}$110 $\alpha$-decay chain. There seems to be 
no new shell stabilizing effects in the considered decays, except for 
$^{10}Be$ decays of $^{269}106$ or $^{273}108$ nuclei, possibly due to the 
known deformed magicity of these superheavy nuclei \cite{hofmann00}. 
Interesting enough, the calculated half-lives for $^{49-51}$Ca decays lie 
far below the present limits of experiments, which go upto $\sim 10^{28} s$ 
\cite{gupta94} for nuclei where enough atoms are available. The closed shell 
effects of a cluster, however, are not yet observed in the exotic cluster 
radioactivity studies. 

\section{Summary of our results}
We have re-investigated the problem of synthesizing superheavy elements via 
the use of $^{48}Ca$ beam, within the quantum mechanical fragmentation 
theory (QMFT) that was first used by Gupta, S\v andulescu and Greiner 
\cite{gupta77b} in 1977. The most probable, specific reaction partners were 
suggested and some of these reactions with $^{48}Ca$ beam on targets of 
$^{232}Th$, $^{238}U$, $^{242,244}Pu$ and $^{248}Cm$ nuclei are recently
used successfully at the JINR Dubna cyclotron U-400. According to the QMFT, 
all these $^{48}Ca$-induced reactions, resulting in different excited 
compound systems, are cold fusion reactions with an "intermediate" amount 
of excitation energy, compared to other cold fusion reactions based on
$Pb$ beams or light nuclei beams such as $Mg$, $Si$, and $S$. The main 
result of the QMFT is that these three types of reactions with different 
excitation energies refer to three different minima in a potential energy 
surface of the same excited compound system having different depths, and are 
due to the shell effects of one or both the reaction partners. The 
re-calculated potential energy surfaces using recent binding energy data 
re-ensure the 1977 result. In fact, the QMFT favours the use of $^{50}Ca$ 
beam for these heavy isotopes and $^{48}Ca$ for the lighter isotopes like 
$^{270}110$, $^{278}112$, $^{284}114$ and $^{290}116$. However, $^{50}Ca$ 
is a radioactive nucleus and hence at present difficult to handle it as a 
beam. Though the experimental physics of radioactive nuclear beams is quite 
different, and at present difficult, the QMFT treat them at par with stable 
nuclei.    

We have also studied the $\alpha$-decay chains of the superheavy nuclei 
$^{277}110$, $^{287,288,289}114$ and $^{292}116$, obtained in ground-states 
after neutron evaporations from the above mentioned excited compound 
systems. The model used is the preformed cluster-decay model (PCM) of Gupta, 
also based on the QMFT. The PCM calculations are then compared with Dubna
experimental data and another recent calculation. Both the calculations 
predict similar trends for $\alpha$-decay half-lives which is due to their 
used Q-values. The experimental $E_{\alpha}$-values (taken as $Q_{\alpha}$-values)
are in general higher and show different trends for atleast some cases and are 
the cause for the missing comparisons between calculations and data by about two 
orders of magnitude. An assuring result of PCM is that $\alpha$-particle 
preformation factors $P_0$ are small, $\sim 10^{-8}$ to $10^{-10}$, required for 
such rare decays. The other calculation uses $P_0=1$. 

Finally, the branching of $\alpha$-decays to other (theoretically possible)
cluster decays are also studied for the $^{277}110$ nucleus. Interesting 
enough, some clusters like $^{49-51}Ca$ have their predicted decay 
half-lives within the present experimental limits for nuclei with enough 
available atoms. Note that here, the not yet observed, shell effects of 
clusters are important. Also, the shell stabilizing effects of deformed and 
weakly deformed Z=108 and 106 nuclei are present in these calculations. 
This means that at present the study of cluster decay of superheavy nuclei 
is mainly of theoretical interest to look for new or already present nuclear 
structure information and perhaps for also pointing out some future 
possibilities with exotic superheavy nuclei. 

\par\noindent
{\bf Acknowledgments:}
RKG, MB and WS are thankful to Volkswagen Stiftung, Germany, for the support 
of this research work under a Collaborative Researach Project between the 
Panjab University and Giessen University. RKG and MB are also thankful to 
the Council of Scientific and Industrial Research (CSIR), New Delhi, for the 
partial support of this research work. 

\vfill\eject

\newpage

{\footnotesize{\scriptsize
\begin{table}[ht]
\noindent
\caption
{The $\alpha$-decay half-lives ($log_{10}T_{\frac{1}{2}}^{\alpha} (s)$) 
and other characteristic quantities for ground-state decays of superheavy
nuclei with Z=110-116 calculated on PCM, and compared with Dubna data
(the $\alpha$-particle energies $E_{\alpha}$ and decay times $\tau$, taken as
$Q_{\alpha}^{expt.}$ and $T_{1\over 2}^{\alpha}$, respectively)
and GLDM calculations. For PCM, the calculations are made at 
$R_a=C_1+C_2=C_t$, using binding energies from Audi-Wapstra and M\"oller et al.
An increase or decrease in $R_a$-value does not improve the PCM results.}
\begin{tabular}{|c|c c c c c|c c|c c c|}\hline \hline
\multicolumn{1}{|c|} {} &\multicolumn{5}{|c|} {PCM} 
&\multicolumn{2}{|c|} {GLDM} &\multicolumn{3}{|c|} {Experiments}\\ \hline
&  case  &$Q_{\alpha}^{case}$ & $P_0$ & P & $log_{10}T_{1\over 2}^{\alpha}$
&$Q_{\alpha}$ & $log_{10}T_{1\over 2}^{\alpha} $ 
&$E_{\alpha}$ & $log_{10}{\tau}$ &chain \\
Parent &        & (MeV) &  & & (s)& (MeV) & (s)&(MeV) & (s) & no. \\ \hline
$^{277}110^*$&cal.&10.696&5.96$\times 10^{-9}$&1.69$\times 10^{-14}$&0.406
&10.89&-4.13&10.31&0.298&1\\ 
&expt.&10.310&5.96$\times 10^{-9}$&3.69$\times 10^{-15}$&1.067&&&&&\\  
$^{273}108$&cal.&9.426&1.14$\times 10^{-9}$&1.71$\times 10^{-16}$&3.121
&9.61&-1.36&escape&2.581&\\ 
$^{269}106$&cal.&7.726&1.53$\times 10^{-10}$&1.93$\times 10^{-20}$&7.939
&7.92&3.59&7.46&3.828&\\
&expt.&7.46&1.53$\times 10^{-10}$&2.56$\times 10^{-21}$&8.816&&&&&\\
\hline   
$^{287}114$&cal.&9.306&3.35$\times 10^{-10}$&3.80$\times 10^{-18}$&5.304
&9.53&0.74&10.29&$0.740^{+0.450}_{-0.196}$&1,2\\
\hline
$^{288}114$&cal.&9.166&3.29$\times 10^{-9}$&1.74$\times 10^{-18}$&2.652
&9.39&1.16&9.87&-0.114&1\\
&&&&&&&&9.80&0.661&2\\
&expt.&9.80&3.29$\times 10^{-9}$&5.74$\times 10^{-17}$&1.133&&&&&\\   
$^{284}112$&cal.&8.696&2.49$\times 10^{-9}$&3.24$\times 10^{-19}$&3.503
&8.89&2.13&9.21&1.013&1\\
&&&&&&&&9.13&1.255&2\\
&expt.&9.13&2.49$\times 10^{-9}$&4.53$\times 10^{-18}$&2.358&&&&&\\
\hline   
$^{289}114$&cal.&8.866&1.44$\times 10^{-10}$&2.84$\times 10^{-19}$&6.797
&9.08&2.16&9.71&1.483&1\\
&expt.&9.71&1.44$\times 10^{-10}$&3.73$\times 10^{-17}$&4.679&&&&&\\   
$^{285}112$&cal.&8.596&1.17$\times 10^{-10}$&1.76$\times 10^{-19}$&7.097
&8.80&2.43&8.67&2.966&1\\
&expt.&8.67&1.17$\times 10^{-10}$&2.84$\times 10^{-19}$&6.889&&&&&\\
$^{281}110$&cal.&8.546&2.09$\times 10^{-10}$&4.47$\times 10^{-19}$&6.439
&8.75&1.95&8.83&1.982&1\\
&expt.&8.83&2.09$\times 10^{-10}$&2.59$\times 10^{-18}$&5.676&&&&&\\
\hline
$^{292}116$&cal.&10.826&7.07$\times 10^{-8}$&2.58$\times 10^{-15}$&-1.850
&11.03&-2.85&10.56&-1.329&1\\
&&&&&&&&10.49&-0.901&2\\
&&&&&&&&10.54&-1.260&3\\
$^{288}114^{\dagger}$&cal.&9.166&3.29$\times 10^{-9}$&1.74$\times 10^{-18}$&2.652
&9.39&1.16&9.81&0.384&1\\
&&&&&&&&9.81&-0.509&2\\
&&&&&&&&9.80&1.040&3\\
$^{284}112^{\dagger}$&cal.&8.696&2.49$\times 10^{-9}$&3.24$\times 10^{-19}$&3.503
&8.89&2.13&$9.09\pm0.46$&1.731&1\\
&&&&&&&&9.15&1.947&2\\
&&&&&&&&9.11&2.184&3\\
\hline\hline
\end{tabular}
\end{table}
\par\noindent
* This chain needs a further careful experimental investigation 
\cite{oganess00}.\\
\par\noindent
${\dagger}$ These isotopes are also observed in an independent decay 
chain of $^{288}114$, shown above.
}}
\vfill\eject

\par\noindent
{\bf Figure Captions}\\
\begin{description}
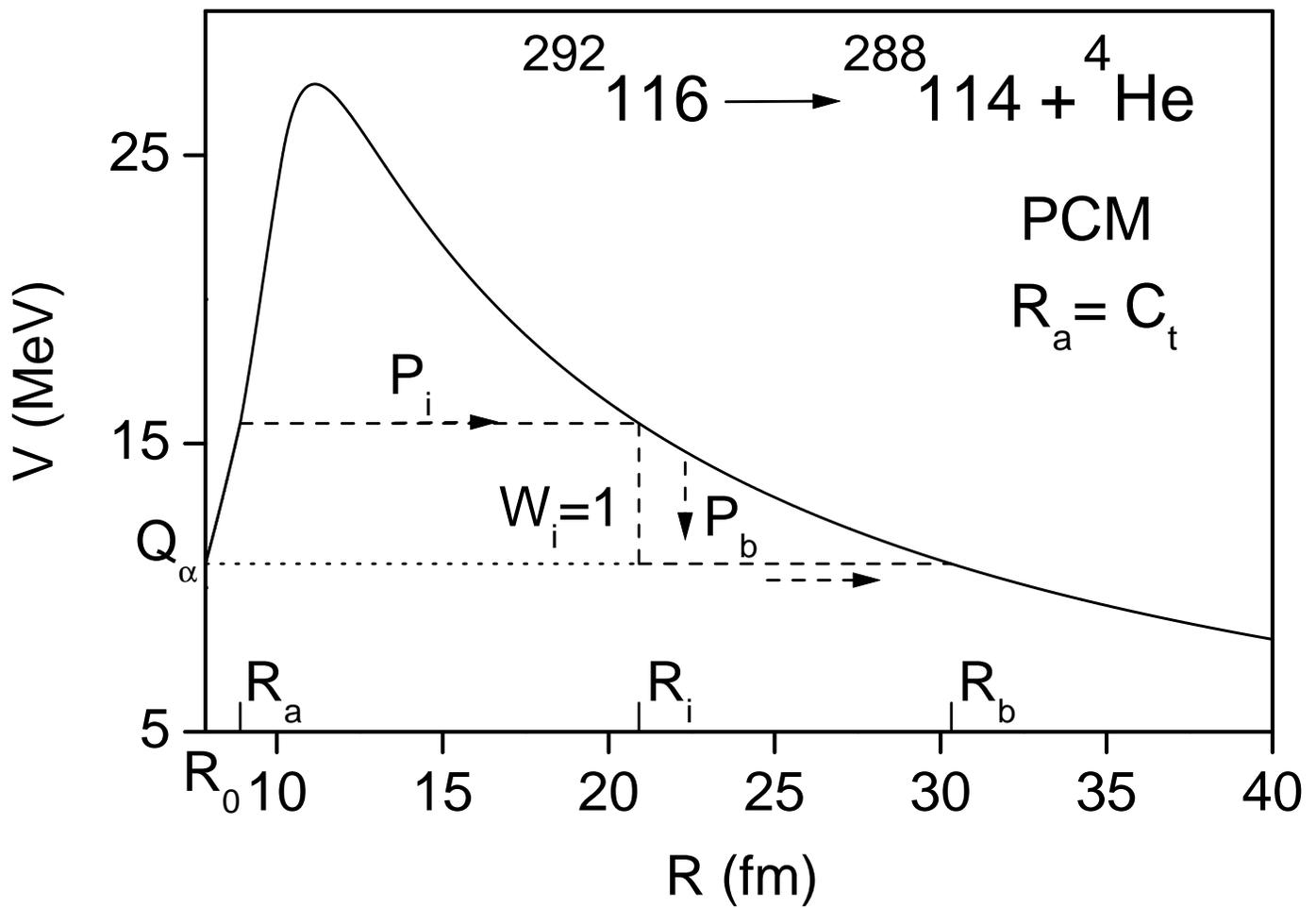
{\item Fig. 1.} The scattering potential for $\alpha$-decay of $^{292}116$ nucleus, 
calculated as the sum of Coulomb and nuclear proximity potential. The 
tunnelling path for PCM is shown, with the first and second turning 
points $R_a$ and $R_b$ also marked.
{\item Fig. 2.} The fragmentation potentials $V(A_2)$, calculated at $R=C_t$, 
for various excited compound systems with Z=110-116. $A_2$ is the mass of 
the lighter nucleus. The heavier nucleus is shown in parentheses for some 
cases only. The ground-state (g.s.) of the compound system is also marked.
{\item Fig. 3.} The calculated formation yields as a function of the mass 
of reaction partners for the compound systems $^{296}116^*$ and $^{292}114^*$ 
at temperatures corresponding to $E_{cm}=$201.1 and 197.2 MeV.
{\item Fig. 4.} The logarithm of $\alpha$-decay half-lives calculated on 
PCM, and compared with the GLDM and experimental data (decay times $\tau$), 
plotted as a function of the parent nucleus mass for various $\alpha$-decay 
chains. The data for all the observed decay chains are shown and the chains are 
identified by numbers. For PCM, a calculation for $Q_{\alpha}^{expt.}$ is also 
shown for $^{288,289}114$ and $^{277}110$ nuclei.  
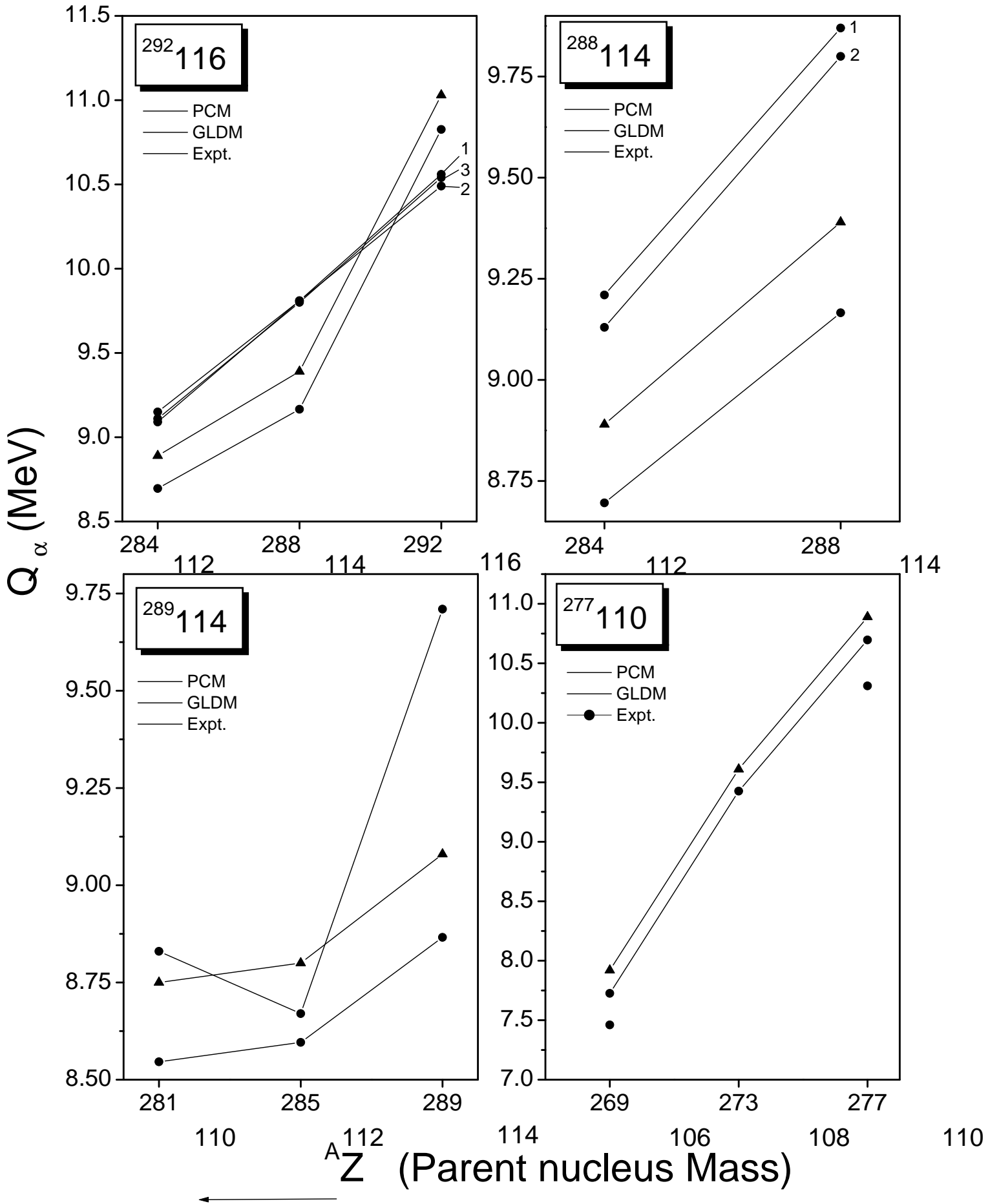
{\item Fig. 5.} The Q-values used in PCM and GLDM, compared with those measured 
($E_{\alpha}$ energies) in $\alpha$-decay chains, plotted as a function of the 
parent nucleus mass. More than one data plotted means more than one observed 
chain, with the same chain identification as in Fig. 4.
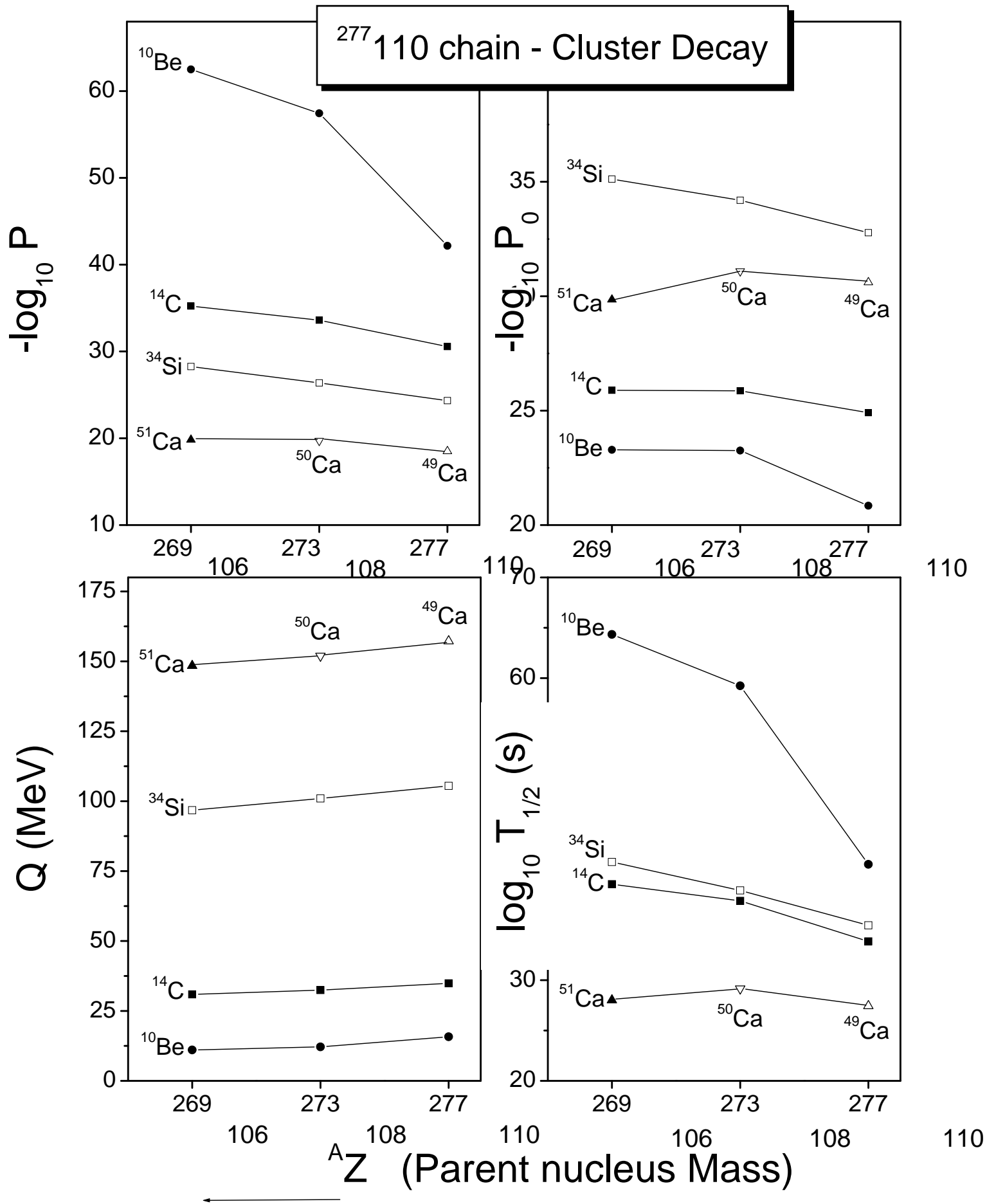
{\item Fig. 6.} The PCM calculated Q-values, penetrability P, preformation 
factors $P_0$, and decay half-lives for some cluster decays of the parents 
of $\alpha$-decay chain for $^{277}$110, plotted as a function of the parent 
nucleus mass.\\
\end{description}

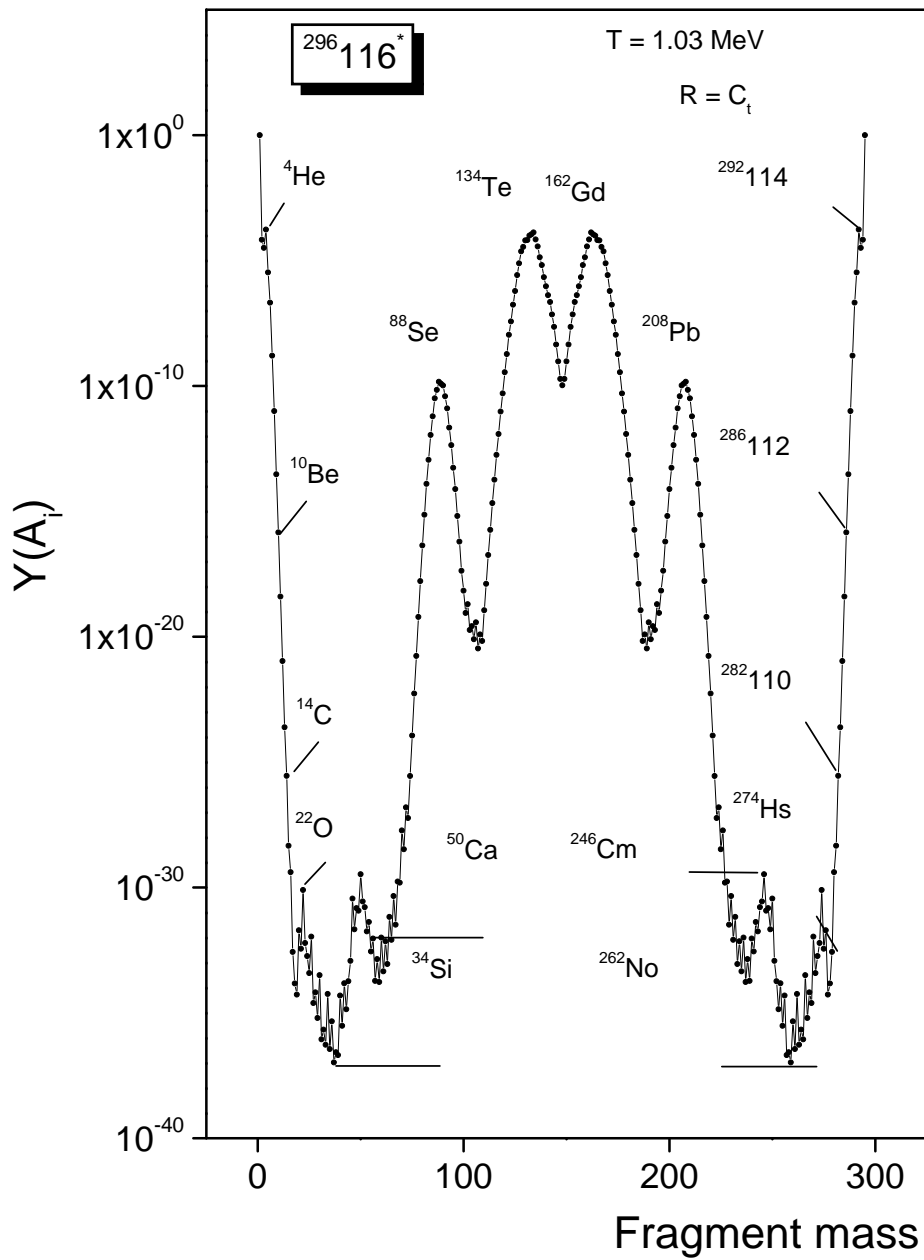
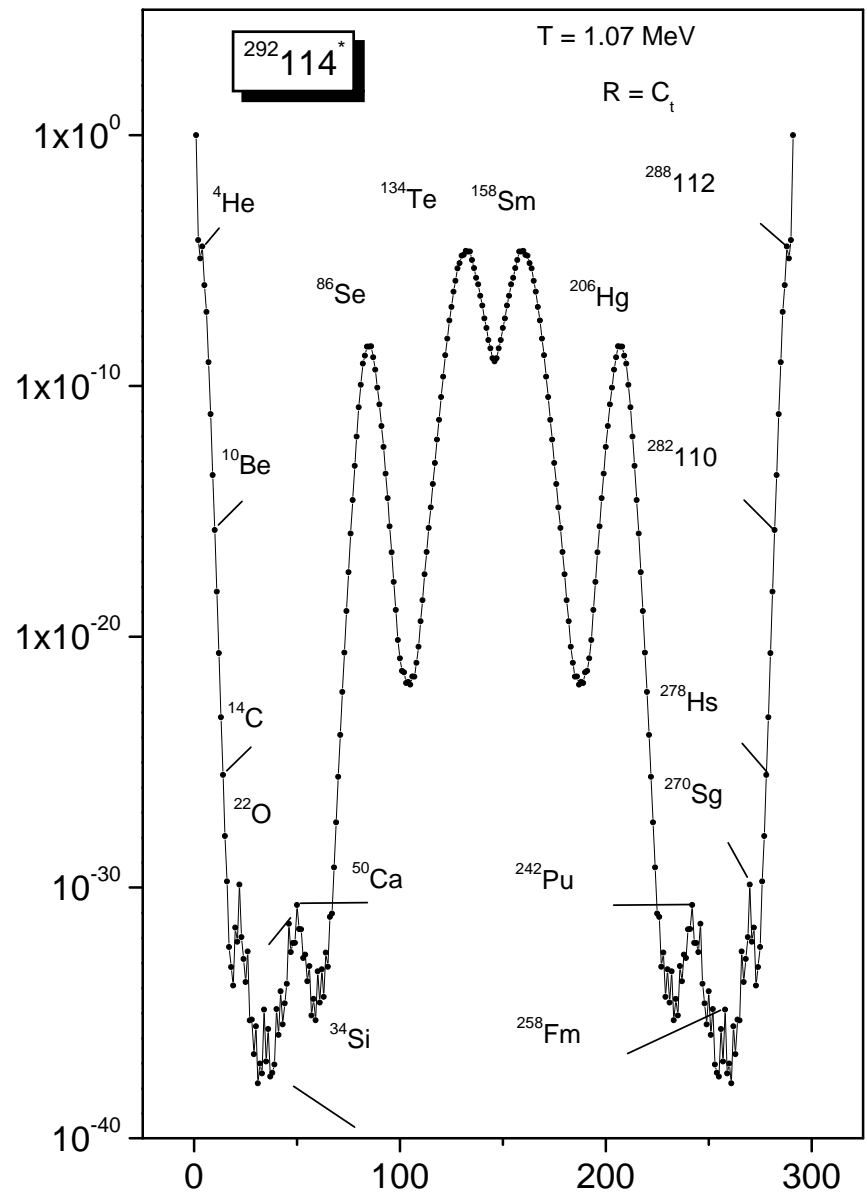
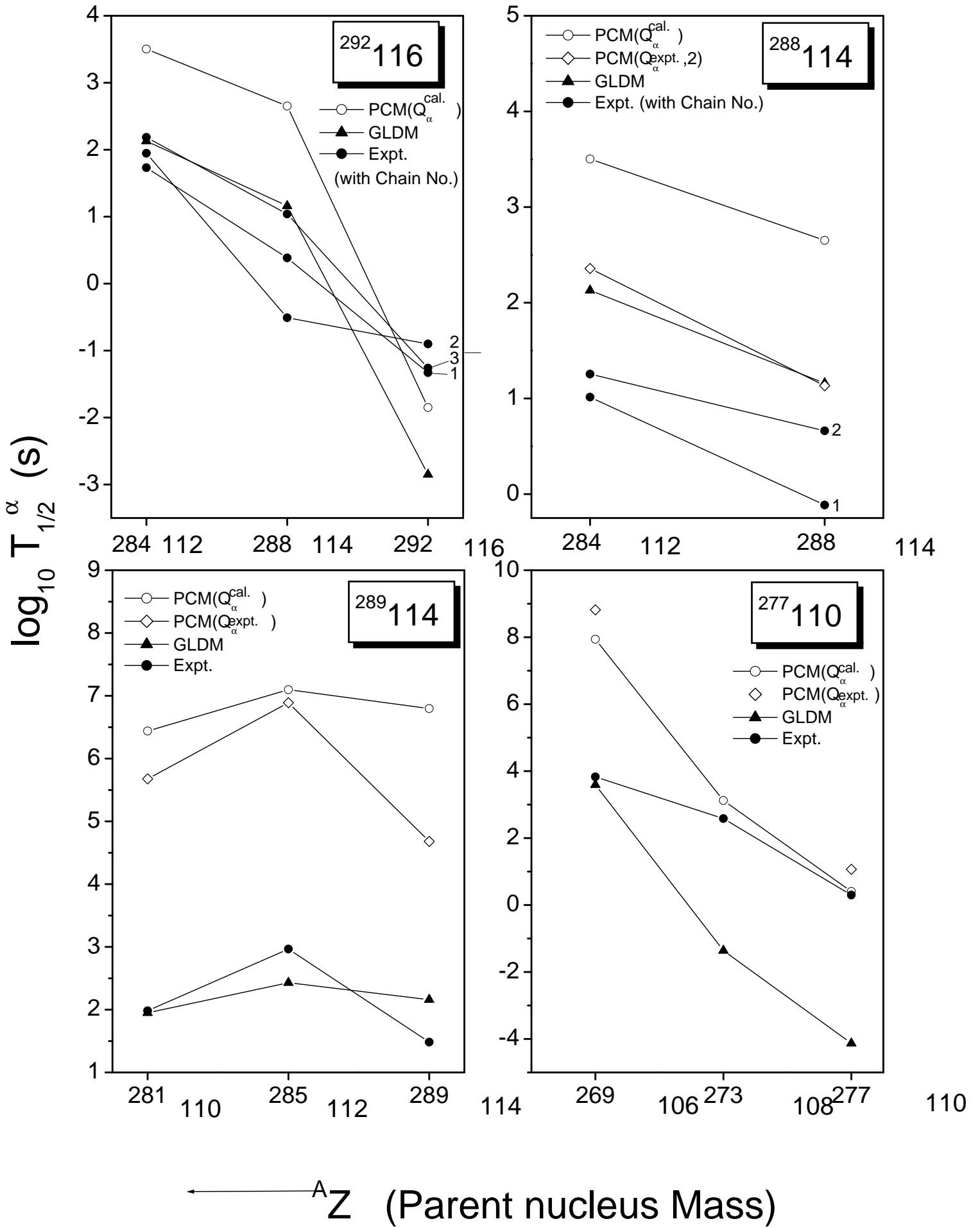
\end{document}